\documentclass[epj]{webofc}
\usepackage[varg]{txfonts}   
\newcommand{\psibar}{\bar{\psi}}
\newcommand{\Nf}{N_{\rm f}}
\newcommand{\rme}{{\rm e}}
\newcommand{\rmO}{{\rm O}}

\wocname{EPJ Web of Conferences}
\woctitle{CONF12}
\begin{document}
\selectlanguage{english}
\title{Probing QCD perturbation theory at high energies \\ with continuum extrapolated lattice data}

\author{Stefan Sint\inst{1}\fnsep\thanks{\email{sint@maths.tcd.ie}}$\,$ for the ALPHA collaboration}

\institute{
School of Mathematics \& Hamilton Mathematics Institute, \\ Trinity College Dublin, 
Dublin 2, Ireland
}

\abstract{%
Precision tests of QCD perturbation theory are not readily available from experimental
data. The main reasons are systematic uncertainties due to the confinement of quarks and gluons,
as well as kinematical constraints which limit the accessible energy scales. We here show how
continuum extrapolated lattice data may overcome such problems and provide excellent
probes of renormalized perturbation theory. This work corresponds to an essential 
step in the ALPHA collaboration's project to determine the $\Lambda$-parameter 
in 3-flavour QCD. 
I explain the basic techniques used in the high 
energy regime, namely the use of mass-independent renormalization schemes for the QCD coupling constant in
a finite Euclidean space time volume. When combined with finite size techniques this allows one 
to iteratively step up the energy scale by factors of 2, thereby quickly covering two orders 
of magnitude in scale. We may then compare perturbation theory 
(with $\beta$-functions available up to 3-loop order) to our non-perturbative data 
for a 1-parameter family of running couplings. We conclude 
that a target precision of 3 percent for the $\Lambda$-parameter requires non-perturbative 
data up to scales where $\alpha_s\approx 0.1$, whereas the apparent precision obtained from applying 
perturbation theory around $\alpha_s \approx 0.2$ can be misleading. 
This should be taken as a general warning to practitioners of QCD perturbation theory.
}
\maketitle
\section{Introduction}
\label{intro}
Lattice QCD is usually thought of as a tool to extract non-perturbative information
about QCD in the hadronic regime. Therefore it is widely believed that lattice QCD is limited to 
the low energy regime where the cutoff (i.e.~the inverse lattice
spacing, $1/a$) sets the limit for accessible scales, typically at a few ${\rm GeV}$.
In this talk I would like to dispel this myth and draw the wider QCD community's attention
to the fact that lattice QCD provides excellent ways to probe perturbation
theory at {\em high} energies and indeed covering a range of energy scales orders of
magnitude apart~(cf.~\cite{Brida:2016flw} and references therein). 
To understand the origin of the above mentioned misconception we first need to
remind ourselves that hadronic physics is done in physically large volumes such that
the linear extent of the volume, $L$, is large in units of the Compton-wave length of the pion, $1/m_\pi$,
which is the lightest particle around. At least for single particle states, the infinite volume limit 
is reached exponentially fast~\cite{Luscher:1985dn} and, depending on the target precision and the particular observables under study, 
it may be sufficient to require $m_\pi L > 4$ for sub-percent errors~\cite{Aoki:2013ldr}. With typical lattice
spacings in the range $0.04$ - $0.1$ ${\rm fm}$, and for pion masses close to physical this implies
lattice sizes up to $L/a=128$. This is at the limit of current technical feasibility 
and there is thus no room to further reduce the lattice spacing without compromising in some other way.

So how can this limitation be overcome? It starts with the simple observation
that perturbation theory can be applied to observables defined in a smaller volume than is required
for hadronic physics. Finite volume effects are then not dominated by pion physics and might
still be under control for the perturbative observables under study. 
One may take this a step further by making the finite volume a part of the definition
of the perturbative observable~\cite{Jansen:1995ck}. In this case there is no need to extrapolate to infinite volume
and one may freely move up and down the energy scale, thus reaching high energies of O(100) ${\rm GeV}$.
The main drawback is that the finiteness of the volume becomes a defining property of the observable, so that
perturbation theory must be done in finite volume, too.  Depending on the chosen boundary conditions 
this may substantially enhance the technical difficulties of perturbation theory.

\section{How to probe the accuracy of perturbation theory}
\label{sect1}

Observables in QCD are defined in terms of renormalized correlation functions 
of gauge invariant composite fields via the QCD path integral,
\begin{equation}
    \left\langle O \right\rangle = {\cal Z}^{-1} \int D[A,\psi,\psibar]\, O[A,\psi,\psibar] \exp\left( -S  \right)\,,
\label{eq:obs1}
\end{equation}
where $A_\mu$, $\psi$ and $\psibar$ denote the gluon, quark and anti-quark fields and we have here assumed
the Euclidean framework. Many such correlation functions have a well-defined perturbative expansions 
in powers of a renormalized coupling, 
$\alpha_s(\mu) = \bar{g}^2(\mu)/4\pi$, 
\begin{equation}
 \left\langle O \right\rangle  = c_0 + c_1 \alpha_s(\mu) + c_2 \alpha_s^2(\mu) +\ldots,
\label{eq:obs2}
\end{equation} 
where the renormalization scale $\mu$ is a priori arbitrary.  However, the perturbative series behaves
best if $\mu$ is chosen close to the relevant physical scales involved in the correlation function, given typically
by particle masses, energies or momenta. Moreover, for the perturbative description to become accurate, 
$\mu$ must be in the ``perturbative regime", $\mu \gg \Lambda$, where the $\Lambda$-parameter is
around a few hundred ${\rm MeV}$ in the $\overline{\rm MS}$ scheme. 
How can a statement about the accuracy of perturbation theory be made more quantitative? 
Besides non-perturbative data, obtained either from experiment or from the lattice, one would like 
to have several orders of the perturbative series available. However, due to the asymptotic nature of the series 
it is probably even more important to vary the size of the expansion parameter, $\alpha_s(\mu)$, 
which is tantamount to varying  the scale $\mu$. 

For observables defined as in eqs.~(\ref{eq:obs1},\ref{eq:obs2}) 
one may normalize the perturbative expansion by defining an ``effective coupling", through
\begin{equation}
 \alpha_O^{}(\mu) =  \dfrac{\left\langle O \right\rangle -c_0}{c_1}  = \alpha_s(\mu) + c_1'\alpha_s^2(\mu) + c_2'\alpha_s^3(\mu)  + \ldots,
\end{equation}
so that its expansion starts with $\alpha_s(\mu)$. In principle, $\alpha_O^{}(\mu)$ 
can be derived from an experimentally measured observable.
However, some practical problems tend to limit the accuracy:
\begin{enumerate}
\item While perturbation theory takes the path integral as starting point, 
its non-perturbative connection to the experimental observable
requires some assumption about the transition from quark and gluon to hadronic degrees of freedom.
The root of the problem is confinement of quarks in hadrons, an inherently 
non-perturbative phenomenon.
\item The scale $\mu$ is usually constrained by the kinematics of the experiment under consideration. 
For example in $\tau$-decays, $\mu$ is essentially determined by the $\tau$-lepton's 
mass~\cite{Pich:2016bdg,Boito:2016oam}.
It is then not possible to vary the energy scale and the only control over the perturbative
expansion is obtained by studying the apparent convergence of the asymptotic series. 
\item For observables defined at energies of a few GeV, 
the dependence of the effective coupling on the charm and bottom quark
masses cannot be ignored so that one uses effective theories 
with $\Nf=3,4,5$ quark flavours, perturbatively matched at the charm and bottom thresholds. 
\end{enumerate}
None of these limitations applies to non-perturbative lattice data in physically small volumes. 
Numerical simulations provide non-perturbative estimates of the Euclidean path integral
which can be directly compared to the perturbative saddle point expansion.
The up, down and strange quark masses can be set to zero and the heavier quarks omitted, thereby
``switching off" the charm and bottom thresholds.  As a result the effective coupling becomes 
a non-perturbatively defined running coupling in a quark mass independent renormalization scheme~\cite{Weinberg:1951ss} 
for 3-flavour QCD, and its scale dependence can be traced over a range of scales 
several orders of magnitude apart.

\section{A family of finite volume couplings}

An attractive class of finite volume couplings can be obtained by
imposing Dirichlet boundary conditions on the fields at Euclidean times $x_0=0$ and $x_0=T$. Due
to its relation with the Schr\"odinger representation in Quantum Field Theory~\cite{Symanzik:1981wd}, the 
path integral in this case is referred to as the Schr\"odinger Functional (SF)~\cite{Luscher:1992an,Sint:1993un}. The SF can be
considered a functional of the boundary values of the gauge field, whereas homogeneous Dirichlet
conditions are imposed on (half the components of) the quark and antiquark fields~\cite{Sint:1993un}.
In a continuum language the spatial components of the gauge potential, $A_\mu$,
are set to abelian, spatially constant boundary fields, $C_k, C'_k$,
\begin{equation}
   A_k(x)\vert_{x_0=0} = C_k,\qquad A_k(x)\vert_{x_0=T} = C'_k\,.
\end{equation}
The boundary fields are chosen 
to depend on 2 real parameters~\cite{Luscher:1993gh}, $\eta,\nu$, corresponding to the 2 abelian generators of SU(3), 
\begin{eqnarray}
C_k^{} &=& \dfrac{i}{L}{\rm diag}\left(\eta-\frac{\pi}{3},\eta\left(\nu-\frac12\right),
                                        -\eta\left(\nu+\frac12\right)+\frac{\pi}{3}\right),\\
C_k'   &=& \dfrac{i}{L}{\rm diag}\left(-\pi-\eta,\eta\left(\nu+\frac12\right)+\frac{\pi}{3},
                                        -\eta\left(\nu-\frac12\right)+\frac{2\pi}{3}\right),
\end{eqnarray}
independently of the direction $k=1,2,3$. For fixed parameters and up to 
gauge equivalence it has been rigorously shown in \cite{Luscher:1992an} that the absolute minimum of the action
corresponds to an abelian spatially constant background field, $B_\mu$,
\begin{equation}
   B_k(x) = C_k + \dfrac{x_0}{T}\left(C_k'-C_k\right),\qquad B_0=0.
\end{equation}
with classical action $S[B]=2(\pi+3\eta)^2/g_0^2$. The effective action is then
unambiguously defined through,
\begin{equation} 
   \rme^{\displaystyle -\Gamma[B]} = \int D[A,\psi,\psibar] \rme^{\displaystyle -S[A,\psi,\psibar]},
\end{equation}
with its perturbative saddle point exansion given as usual by
\begin{equation}
 \Gamma[B] = \frac{1}{g_0^2}\Gamma_0[B] + \Gamma_1[B] + \rmO(g_0^2), \qquad \Gamma_0[B] = g_0^2 S[B]\,.
\end{equation}
A family of couplings in the SF scheme is thus obtained by defining~\cite{Sint:2012ae,Brida:2016flw}
\begin{equation}
   \dfrac{1}{\bar{g}_\nu^2(L)} = \left.\dfrac{\partial_\eta \Gamma[B]}{\partial_\eta \Gamma_0[B]}\right\vert_{\eta=0}
   = \dfrac{\left.\left\langle \partial_\eta S \right\rangle\right\vert_{\eta=0}}{12\pi}
   = \dfrac{1}{\bar{g}^2(L)} - \nu\times \bar{v}(L) 
   \label{eq:gnudef}
\end{equation}
Note that the $\eta$-derivative produces an expectation value of an observable 
which can be measured in a numerical simulation. While we set $\eta=0$, 
the dependence on the parameter $\nu$ is completely explicit, 
so that a calculation at $\nu=0$ yields the full 1-parameter
family of couplings, $\bar{g}^2_\nu$, in terms of 
$\bar{g}^2 = \bar{g}^2_{\nu=0}$ and a second observable, $\bar{v}$.
The $\eta$-derivative can be interpreted as a variation of the background field,
and the SF couplings are thus defined by the response of the system 
to a change of such a colour electric field. Finally, the quark masses are set to zero and
any remaining dimensionful quantities  such as the Euclidean time extent, $T$, or the 
strength of the background field are scaled proportionally to $L$. The couplings
thus depend on a single scale which is naturally identified with the
renormalization scale, i.e.~$\mu=1/L$.

\section{Properties of SF schemes}

In perturbation theory the SF coupling \cite{Luscher:1993gh,Sint:1995ch} 
has been matched to the $\overline{\rm MS}$ coupling up to
2-loop order~\cite{Bode:1999sm,Luscher:1995nr,Christou:1998wk,Christou:1998ws}
and therefore its 3-loop $\beta$-function can be inferred from 
the $\overline{MS}$-scheme~\cite{MS:4loop1,Czakon:2004bu}. In our conventions we have
\begin{equation}
   \beta(\bar{g}) = -L \dfrac{\partial \bar{g}(L)}{\partial L},\qquad \beta(g) = -b_0 g^3 - b_1 g^5 + \ldots,
\end{equation}
with universal coefficients,
\begin{equation}
   b_0 = \left(11 -\tfrac23 \Nf\right)/(4\pi)^{2}, \qquad  
   b_1 = \left(102 -\tfrac{38}{3}\Nf\right)/(4\pi)^{4}\,,
\end{equation}
and the 3-loop coefficient(s),
\begin{equation}
   b_{2,\nu} = \left(-0.06(3) -\nu\times 1.26\right)/(4\pi)^3\,.
\end{equation}
For tests of perturbation theory the definition of our observables 
in a finite Euclidean space-time volume represents a real advantage. 
In particular, the infrared cutoff by the finite
volume prevents any renormalon issues. The fact that the minimum action configuration is unique makes 
the saddle point expansion straightforward. However, as is always the case
with asymptotic expansions, exponentially small corrections to the series 
are neglected.  Their size depends on the choice of observable and the value
of the coupling. In our case such terms originate from secondary minima of the classical action and
one would expect such contributions to be suppressed by
$ \exp(-\Delta S)$ with $\Delta S$ the difference between
the classical action taken at a secondary minimum and at the absolute minimum, respectively.
We have investigated this issue~\cite{SFcoupinpreparation} and found the nearest stable stationary
point of the action corresponds to $g^2 \Delta S = 10\pi^2/3$. Hence, for the 
range of couplings used in our work such contributions are completely negligible.

\section{Accuracy of perturbation theory in terms of the $\Lambda$-parameter}

In order to define a target accuracy for the comparison 
with perturbation theory it is useful to refer to the $\Lambda$-parameter,
which, in a mass-independent renormalization scheme, is defined as an 
exact solution to the Callan-Symanzik equation, viz.
\begin{equation}
   \left( \mu\dfrac{\partial}{\partial \mu} + \beta(g) \dfrac{\partial}{\partial g}\right) \Lambda = 0\,.
\end{equation}
Non-perturbatively defined couplings imply the non-perturbative definition 
of the corresponding $\beta$-function
and one obtains the exact solution ($\mu=1/L$),
\begin{equation}
    L\Lambda = \varphi\left(\bar{g}(L)\right),\qquad   
    \varphi\left(\bar{g}\right) = \big[{b_0\bar{g}^2}\big]^{-\frac{b_1}{2b_0^2}}\,\rme^{-\frac{1}{{2b_0\bar{g}^2}}}
                       \exp\bigg\{\!\!-\!\!\int_{0}^{\bar{g}^{\vphantom{A}}}\! dg
                       \bigg[\dfrac{1}{\beta(g)}+\dfrac{1}{b_0g^3} -\dfrac{b_1}{b_0^2g}\bigg]\bigg\}\,.
\end{equation}
The coupling, its $\beta$-function, and thus $\varphi$ and $\Lambda$ depend on the renormalization scheme. 
However, the scheme dependence of the $\Lambda$-parameter is almost trivial: assuming 2 schemes $X$ and $Y$ the 
matching of the respective couplings to one-loop order entails the {\em exact} relation,
\begin{equation}
    g^2_{\rm X}(\mu) = g^2_{\rm Y}(\mu) + c_{\rm XY}^{} g^4_{\rm Y}(\mu) + ...
    \quad \Rightarrow \quad \dfrac{\Lambda_{\rm X}}{\Lambda_{\rm Y}} = \mathrm{e}^{c_{\rm XY}^{}/2b_0}\,,
\end{equation}
so that we may use $\Lambda = \Lambda_{\text{SF,$\nu=0$}}$ as reference.
Introducing a reference scale $1/L_0$ through,
\begin{equation}
   \bar{g}^2(L_0) = 2.012 \quad \Rightarrow\quad
   \dfrac{1}{\bar{g}_\nu^2(L_0)} = \dfrac{1}{2.012} - \nu\times \underbrace{\bar{v}(L_0)}_\text{eq.~(\ref{eq:vbarL0})}\,,
  \label{eq:nustart}
\end{equation}
we consider the reference quantity,
\begin{equation}
  L_0\Lambda = \underbrace{ \Lambda/\Lambda_\nu }_{\exp(-\nu\times1.25516)} \quad \times \quad \underbrace{L_0/L}_{\text{non-perturbative}}
  \quad \times \quad \underbrace{\varphi_\nu\left(\bar{g}_\nu(L)\right)}_{\text{perturbative}}\,.
  \label{eq:L0Lambda}
\end{equation}
Hence, if we use non-perturbative results for the running coupling in the range 
$1/L_0 \leq \mu \leq 1/L$, and apply perturbation theory for $\mu > 1/L$ 
by replacing $\beta_\nu(g)$ by its 3-loop approximation, $\beta_{\nu,\text{3-loop}}(g)$, 
\begin{equation}
\varphi_\nu\left(\bar{g}_\nu(L)\right) \propto 
 \exp\bigg\{\!\!-\!\!\int_{0}^{\bar{g}_\nu^{\vphantom{A}}(L)}\! dg 
  \bigg[\dfrac{1}{ \beta_{\nu,\text{3-loop}}(g)}+\dfrac{1}{b_0g^3} -\dfrac{b_1}{b_0^2g}\bigg]\bigg\}, \qquad
 \beta_{\nu,\text{3-loop}}(g) = -b_0 g^3 - b_1 g^5 - b_{2,\nu}\, g^7, \\
 \end{equation}
we should find that $L_0\Lambda$ is independent of the choice of $L$ and $\nu$, up
to perturbative errors at the scale $\mu=1/L$. Moreover, we have some idea of the target precision 
for the $\Lambda$-parameter: if we aim for, say 0.5 percent accuracy for $\alpha_s(m_Z)$ then
the 3-flavour $\Lambda$-parameter should be determined to better than 3 percent accuracy~\cite{Agashe:2014kda}.
Imposing this criterion on $L_0\Lambda$ thus gives us a handle to assess the accuracy of
perturbation theory when applied at scales $\mu > 1/L$.
We just have to explain how to trace the non-perturbative running of the SF coupling
from $1/L_0$ to $1/L$, which should be a range covering a couple of orders of magnitude.

\section{Non-perturbative running in steps and determination of $L_0/L$}

The starting point is the so-called step-scaling function~\cite{Luscher:1991wu},
\begin{equation}
 \sigma(u) = \left.\bar{g}^2(2L)\right\vert_{u=\bar{g}^2(L)},
\end{equation}
which determines how the coupling at scale $1/(2L)$ depends on the coupling at scale $1/L$.
Hence one considers scales which are separated by a factor 2 rather 
than infinitesimally as for the $\beta$-function. The relation to the latter
is defined by
\begin{equation}
 \int_{\sqrt{u}}^{\sqrt{\sigma(u)}} \dfrac{dg}{\beta(g)} = -\ln 2\,.
\end{equation}
The step-scaling function $\sigma(u)$ is defined in the continuum limit.
In order to obtain this function at a fixed value of its argument, 
one measures the coupling on pairs of lattices with extent $L/a$ and $2L/a$ 
in each direction. Since all quark masses are set to zero, the only parameter
to be tuned is the bare coupling which is equivalent to the lattice spacing $a$.
\begin{figure}[t]
\centering
\sidecaption
\includegraphics[width=7cm,clip]{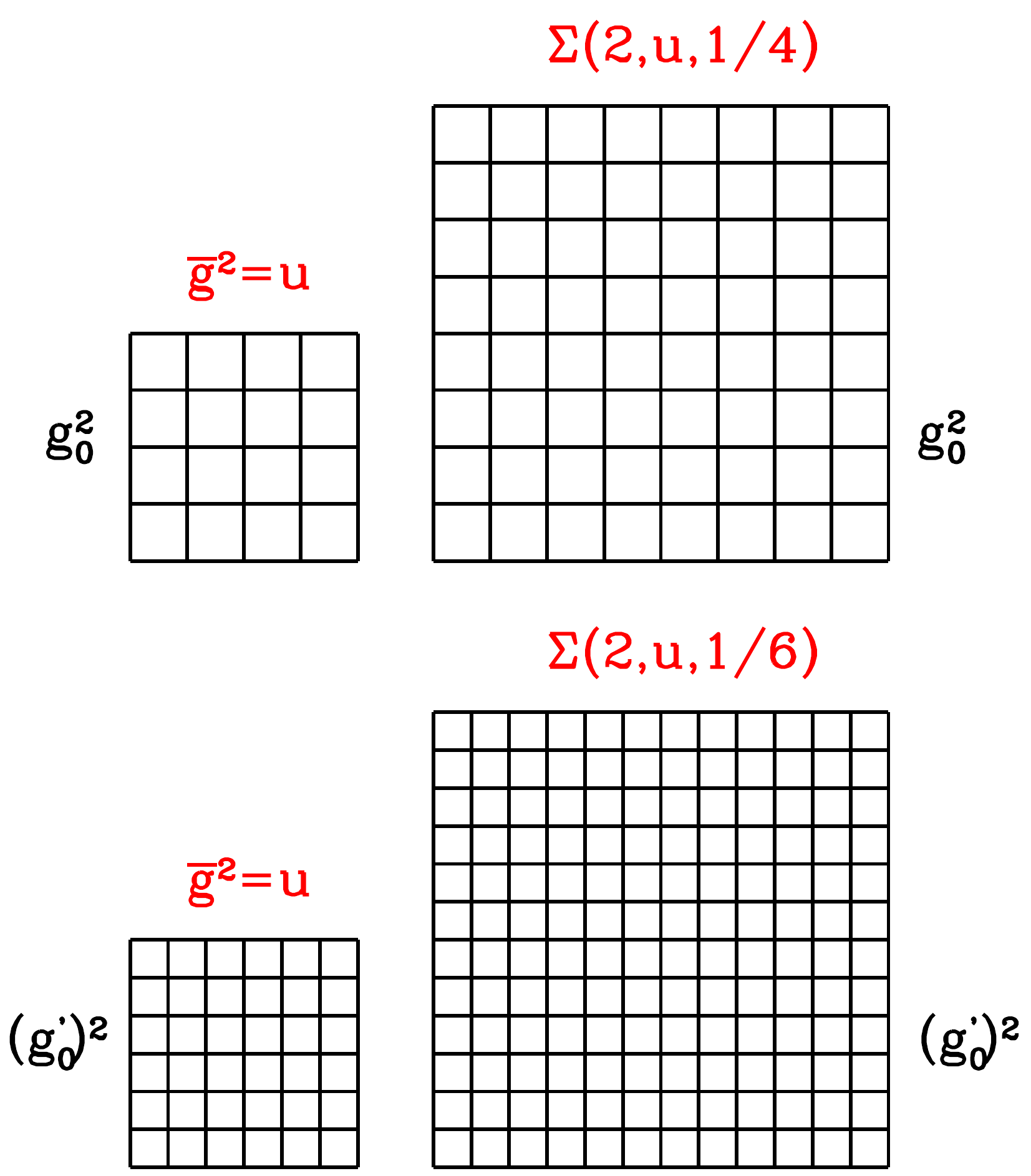}
\caption{Illustration of the computation of the step scaling function, 
$\Sigma(u,a/L)$, at a fixed value of the coupling $u=\bar{g}^2(L)$ 
and for 2 lattice resolutions $L/a=4,6$ (cf.~text).}
\label{fig-2}       
\end{figure}
Hence if we start with a lattice size $L/a=4$, measure $u=\bar{g}^2(L)$ in a numerical
simulation and then measure $\bar{g}^2(2L)$ by doubling the lattice size to $2L/a$
(at the same value of the bare coupling $g_0$), we obtain a first approximant, $\Sigma(u,1/4)$,
for $\sigma(u)$, where the second argument is the resolution $a/L$.
We now want to keep $L$ constant in physical units but choose an $L/a=6$ lattice. 
This means we have to tune the bare coupling to another value $g_0'$ such 
that the previous $u$-value is matched
(which implicitly fixes $L$). Then, at the same $g_0'$ one doubles the lattice size
and measures the SF coupling to obtain $\Sigma(u,1/6)$. Continuing the procedure
one may take the limit,
\begin{equation}
   \sigma(u) = \lim_{a/L\rightarrow 0} \Sigma(u,a/L)\,,
\end{equation}
at the given value of $u$. Repeating the same procedure for a range of $u$-values
one obtains the function $\sigma(u)$ for this range with a certain error, due to
both statistics and systematic effects from the continuum extrapolation.
Once $\sigma(u)$ is available for a range of values $u \in [u_{\text{min}},u_0]$ 
one may iteratively step up the energy scale:
\begin{equation}
  u_0=\bar{g}^2(L_0),  \quad u_{n} = \sigma(u_{n+1}) = \bar{g}^2(L_n) = \bar{g}^2(2^{-n}L_0), \quad n=0,1,...
\end{equation}
In particular, by construction the scale ratios are $L_0/L_n = 2^n$, 
where $n$ is the number of steps.
\begin{figure}[t]
\centering
\sidecaption
\includegraphics[width=7cm,clip]{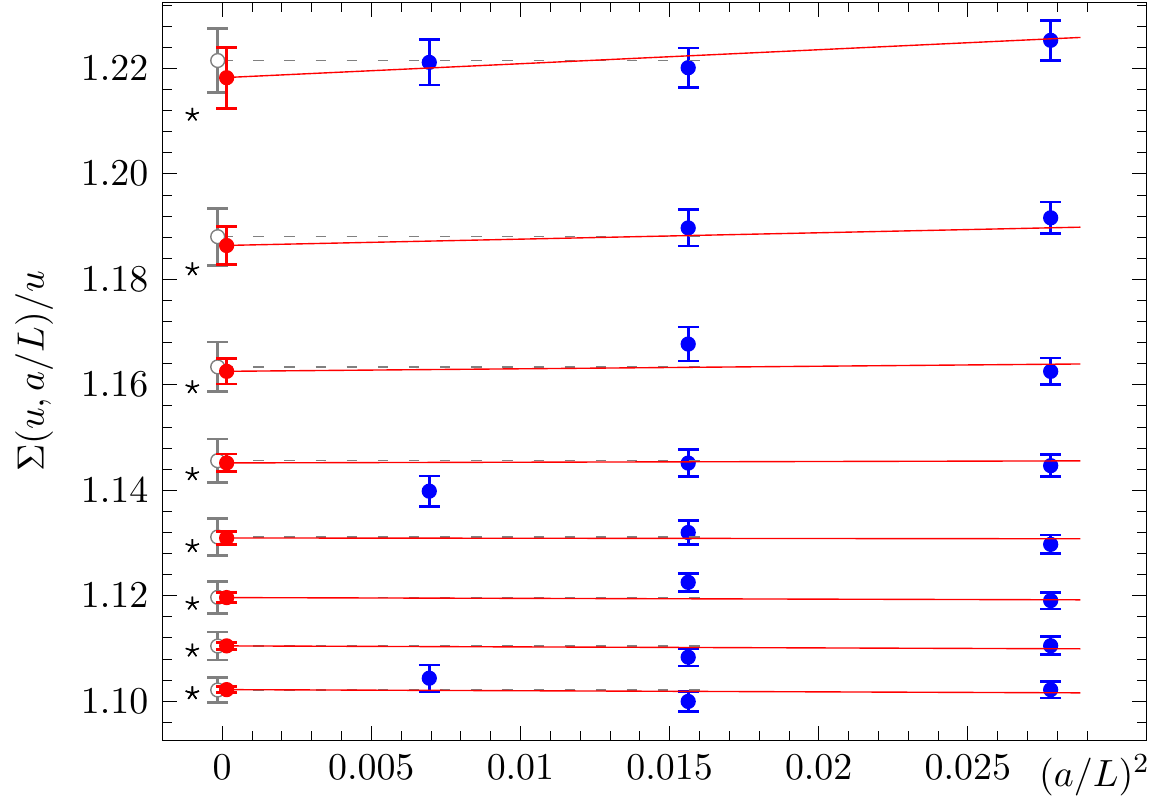}
\caption{Continuum extrapolation of the step scaling function. Some data points
have been slightly shifted in order to keep constant $u$, which is
not required for the global fit. The leftmost points are the continuum values, whereas
the stars are obtained from perturbative scale evolution using the 3-loop $\beta$-function.}
\label{fig:CLssf}
\end{figure}

\section{Continuum extrapolation of  $\Sigma(u,a/L)$}

The lattice step-scaling function $\Sigma(u,a/L)$ is expected to be a smooth function
of $u$. Lattice effects are, up to slowly varying logarithmic terms, 
polynomial in $a/L$. Moreover, terms linear in $a/L$ are removed in
the bulk by the use of the non-perturbatively O($a$) improved action~\cite{Yamada:2004ja},
and highly suppressed at the time boundaries by using perturbative
estimates of the counterterm coefficients. As a safeguard against
any residual O($a$) effects we treat a variation of the 2 counterterm
coefficients as a systematic effect and propagate it to the data.
Rather than extrapolating the step-scaling function separately for individual values of $u$ 
it is more practical to use a global fit ansatz. A typical example is
\begin{equation}
\Sigma(u,a/L) = u + s_0 u^2 + s_1 u^3 
+  { c_1} u^4 + { c_2} u^5 
+  {\rho_1} u^4 \dfrac{a^2}{L^2} + {\rho_2} u^5 \dfrac{a^2}{L^2}
\end{equation}
with $s_0$, $s_1$ fixed to perturbative values,
\begin{equation}
  s_0 = 2b_0\ln 2,\quad  s_1 = s_0^2 + 2b_1\ln 2\,.
\end{equation} 
Our data for $\Sigma(u,a/L)$  has $L/a=4,6,8$ and, for a couple of $u$-values we have $L/a=12$. 
As a safeguard we omit the coarsest lattices with $L/a=4$ and fit the 19 data points to the above fit ansatz with
4 free parameters, $c_1,c_2,\rho_1,\rho_2$.  Fig.~\ref{fig:CLssf} shows the data
together with the fit function. The $\chi^2/\text{d.o.f.}\approx 1$ for this
and a variety of different fit ans\"atze indicates that 
we have a good control over the continuum limit. The continuum SSF is
then represented by $\sigma(u) =  u+ s_0 u^2 + s_1 u^3 
+ c_1 u^4 + c_2 u^5$, in the fit range $[1.02,2.02]$ and with numerical values
for $c_1,c_2$, together with their errors and correlation.
The continuum step-scaling function can be compared with earlier results in~\cite{Aoki:2009tf}.
where an attempt is made to reach larger couplings so as to match hadronic scales.
In the high energy regime we have significantly improved on the precision both in terms
of statistical and systematic errors, for instance through a very precise tuning to 
zero quark mass~\cite{PatrickTomaszinpreparation}.

\section{Results}

\begin{figure}[h]
\centering\sidecaption
\includegraphics[width=8.5cm,clip]{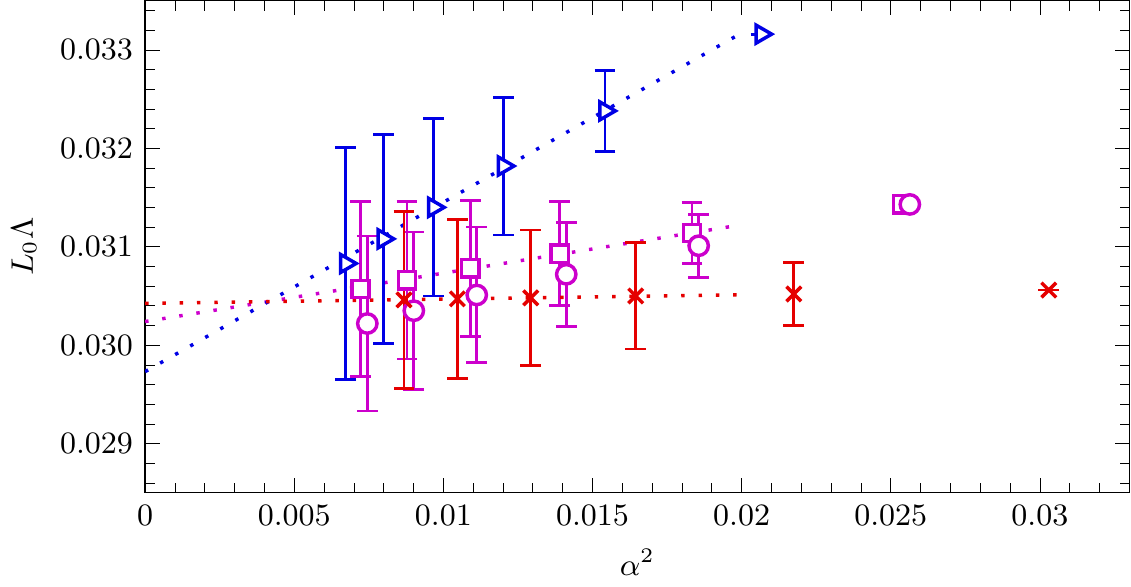}
\caption{The extraction of the $\Lambda$-parameter using perturbation theory at various values of $\alpha_s$, plotted vs. $\alpha_s^2$. 
The data points are, from top to bottom, for $\nu=-0.5,0,0.3$ and, from right to left, for $n=0,1,\ldots,5$ steps by a factor 2 in scale.
Nice agreement is observed around $\alpha_s\approx 0.1$.}
\label{fig:LambdaL0}       
\end{figure}
\begin{figure}[h]
\centering
\sidecaption
\includegraphics[width=8.5cm,clip]{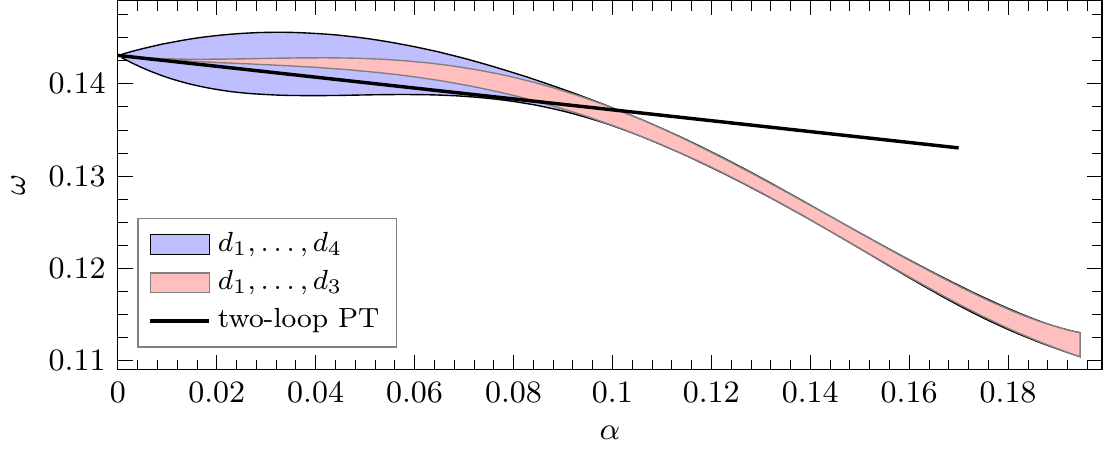}
\caption{The observable, $\omega(u)=\bar{v}(L)$, plotted vs. $\alpha_s$, together with the
fits eqs.~(\ref{eq:fit1},\ref{eq:fit2}) and the two-loop result (cf.~text).}
\label{fig:vbar}       
\end{figure}

The step scaling functions have been analysed for a number of $\nu$-values of O(1).
Using eq.~(\ref{eq:L0Lambda}) for $L_0\Lambda$ we use non-perturbative running
between $L_0$ and $L_n=2^{-n}L_0$, with $n=0,1,\ldots,5$. In physical units
$1/L_0$ is about 4 GeV, so that we cover a range from 4 to 128 GeV.
For a given $n$ we then integrate the integral in the exponent
using the $\beta$-function to 3-loop order. 
In fig.~\ref{fig:LambdaL0} the results obtained at various $n$ and
the 3 values $\nu=0,0.3,-0.5$ are plotted vs.~$\alpha_s^2$, which
is the order of the neglected terms. Up to such terms all the data points should agree within errors.
The expected asymptotic behaviour is indeed observed.  
We see that for $\nu=0.3$ the slope in $\alpha_s^2$ is essentially zero,
whereas it is rather large at $\nu=-0.5$. From fig.~\ref{fig:LambdaL0} and a variety of
further fits not shown here we conclude that all results agree around $\alpha=0.1$ at the 
3 percent level or better,
\begin{equation}
   L_0\Lambda = 0.0303(8) \quad \Leftrightarrow \quad 
   L_0\Lambda_{\overline{\rm MS}}^{\Nf=3} = 0.0791(21)\,.
\label{eq:LambdaL0}
\end{equation}
While for $\nu=0.3$  this result could be inferred from larger values of $\alpha_s$, this is clearly
not the case for $\nu=-0.5$. To further assess the accuracy of perturbation theory
it is instructive to directly look at the second observable
\begin{equation}
  \bar{v}(L) = \omega(u)\vert_{u=\bar{g}^2(L)}\,,
\end{equation}
which allows us to study the couplings for all $\nu$-values (\ref{eq:gnudef})
We have extrapolated the non-perturbative data to the continuum
limit using similar global fits as for the step-scaling functions.
However, here these fits are more constrained as there is no doubling of 
the lattice size involved and lattice sizes thus range from $L/a=6$ to $L/a=24$.
Two resulting fit functions in the continuum limit of the form 
\begin{eqnarray}
   \omega(u)\vert_{\text{fit 1}} &=& v_1+v_2u+d_1u^2+d_2u^3+d_3u^4\,, \label{eq:fit1}\\
   \omega(u)\vert_{\text{fit 2}} &=& v_1+d_1u+d_2u^2+d_3u^3+d_4u^4\,, \label{eq:fit2}
\end{eqnarray}
with 3 and 4 fit parameters $d_k$, $k=1,\ldots$, respectively, are shown with their
error bands in fig.~\ref{fig:vbar}. Both fits agree perfectly well in the whole range
of the available non-perturbative data. The continuum result at $\bar{g}^2(L_0)=2.012$,
\begin{equation}
 \bar{v}(L_0) = \omega(2.012) = 0.1199(10),
 \label{eq:vbarL0}
\end{equation}
is obtained from these fits and defines the starting values for the step-scaling
procedure for $\nu\ne 0$ [cf.~eq.~(\ref{eq:nustart})].
Fig.~\ref{fig:vbar} also shows the known 2-loop result
\begin{eqnarray}
  \omega(u) = \left.\bar{v} \right|_{\bar{g}^2(1/L)=u,m=0} 
  = v_1 + v_2 u + \rmO(u^2) \,,
  \label{eq:omega}
\end{eqnarray}
where the coefficients $v_1,\,v_2$ can be found in~\cite{DellaMorte:2004bc}).
The non-perturbative data clearly breaks away from two-loop
perturbation theory at larger couplings. To quantify this 
deviation we choose the value $\alpha_s=0.19$
and measure an effective 3-loop coefficient as follows
\begin{eqnarray}
   (\omega(\bar{g}^2) - v_1 - v_2 \bar{g}^2 )/v_1 = -3.7(2) \,\alpha_s^2\,.
   \label{eq:vbar3eff}
\end{eqnarray}
Indeed this effective coefficient seems too large for perturbation theory
to be trustworthy at this value of the coupling.
We come to the conclusion that  $\alpha\approx 0.1$
needs to be reached non-perturbatively for perturbation theory to become
as accurate as we have required here.

\section{Conclusions}

We have pointed out that lattice observables can be defined at high energies 
if one gives up the requirement that volumes should be large enough to fit hadronic states. 
By defining observables in a finite volume it is possible 
to obtain non-perturbative precision data over a wide range of scales. Moreover,
the heavy quark thresholds for charm and bottom can be ``switched off" on the lattice,
thereby removing an important source of systematic errors.
The main drawback is that perturbation theory must match this situation
and take the finite volume into account. This implies some technical difficulties 
which depend on  all the details of the chosen set-up. With the SF scheme chosen here there exists a
2-loop calculation matching the SF couplings to the $\overline{\rm MS}$-scheme and thus
the 3-loop $\beta$-function is known for a whole 1-parameter family of SF couplings.
This provides excellent opportunities to test the accuracy of perturbation theory.
As it turns out, a precision of 3 percent for the $\Lambda$-parameter can be quoted with confidence
if perturbation theory is restricted to couplings around $\alpha_s\approx 0.1$ or smaller. 
However, at $\alpha_s \approx 0.2$ and to this level of accuracy
there is much less confidence in perturbation theory and some luck is required when 
choosing a scheme. 

Our result for the $\Lambda$-parameter, eq.~(\ref{eq:LambdaL0}), is an essential step in the ALPHA collaboration's project
to determine the $\Lambda$-parameter in 3-flavour QCD in units of a hadronic scale such
as the kaon and pion decay constants, $F_{K,\pi}$. These are determined in large volumes~\cite{Bruno:2016plf}
on gauge configurations produced through the CLS effort~\cite{Bruno:2014jqa}.
Preliminary results have been presented by M.~Dalla Brida at this conference and by
R.~Sommer in \cite{Bruno:2016gvs}. An estimate of $\alpha_s(m_Z)$ is given there, assuming that the standard
perturbative treatment of the charm and bottom quark thresholds is reliable.
While we have focussed here on the $\Lambda$-parameter and running
couplings, an analogous study can be made for the running quark masses and first
results have been presented by P.~Fritzsch at this conference~\cite{Campos:2016vxh}.

\begin{center}
{\bf Acknowledgments}
\end{center}
This work was done as part of the ALPHA collaboration research programme. 
I would like to thank the members of the ALPHA collaboration and particularly my 
co-authors of ref.~\cite{Brida:2016flw} for the enjoyable collaboration on this project and for comments on the manuscript.
Computer resources by the computer centres at HLRN (bep00040) and NIC at DESY, Zeuthen, as well as
financial support by SFI under grant 11/RFP/PHY3218 are gratefully acknowledged. 
\bibliography{main}  
\end{document}